\documentclass[fleqn,12pt,twoside]{article}
\usepackage[headings]{espcrc1}
\def\be{\begin{equation}}
\def\ee{\end{equation}}
\def\bea{\begin{eqnarray}}
\def\eea{\end{eqnarray}}
\def\beq{\begin{equation}}
\def\eeq{\end{equation}}

\def\mk{{\mathbf k}}%

\readRCS
$Id: espcrc1.tex,v 1.2 2004/02/24 11:22:11 spepping Exp $
\usepackage{graphicx}
\usepackage[figuresright]{rotating}

\newcommand{\AmS}{{\protect\the\textfont2
  A\kern-.1667em\lower.5ex\hbox{M}\kern-.125emS}}

\title{Survival of Back-to-Back Correlations for Finite Expanding Fireballs}
\author{\bigskip Sandra S. Padula\medskip \address[MCSD]{Inst. de F\'{\i}sica Te\'orica,
UNESP~-~Rua Pamplona 145, 01405-900 S\~ao Paulo, SP - Brazil}%
        \thanks{SSP would like to express her gratitude to the Organizing Committee
        of Quark Matter 2005 and to 
        FAPESP (Proc. N. 05/52190-1 and 2004/10619-9), S\~ao Paulo, Brazil, 
for their complementary financial support. 
This work was partially supported by OTKA T038406 and T049466.},
        Y. Hama\address{Instituto de F\'{\i}sica, USP, Caixa Postal 66318, 05389-970 S\~{a}o Paulo, SP -
        Brazil},
        G. Krein\addressmark[MCSD],
        P. K. Panda\address{Depto. de F\'\i sica-CFM, UFSC - C. P. 476, 88040-900 Florian\'opolis, SC - Brazil}
        and
        T. Cs\"org\H o\address{MTA KFKI RMKI, H - 1525 Budapest 114, POBox 49,
        Hungary}}

\runtitle{Survival of Back-to-Back Correlations for Finite
Expanding Fireballs} \runauthor{Sandra S. Padula et al.}

\begin{document}

\maketitle

\begin{abstract}

\end{abstract}


\bigskip In the late 1990's, {\sl Back-to-Back Correlations} ({\sl
BBC}) of boson-antiboson pairs were predict to exist if the
particles masses were modified in the hot and dense
medium\cite{acg99}, expected to be formed in high energy
nucleus-nucleus collisions. The BBC are related to in-medium
mass-modification and squeezing of the quanta involved. Not much
longer after that, it was also shown that an analogous BBC existed
between fermion-antifermion pairs with medium-modified
masses\cite{pchkp01}. A similar formalism is applicable to both
BBC cases, related to the Bogoliubov-Valatin transformations of
in-medium and asymptotic operators. Both the bosonic (bBBC) and
the fermionic (fBBC) Back-to-Back Correlations are positive and
have unlimited magnitude, thus differing from the
identical-particle correlations, also known as HBT (Hanbury Brown
\& Twiss) correlations, which are limited for both cases, being
negative in the fermionic sector. BBC were expected to be
significant for $p_T < 2$ GeV/c. Nevertheless, already in the
Ref.\cite{acg99}, it was shown that, if the emission process is
not sudden, even a short duration of particle emission
significantly suppresses the BBC magnitude. On the other hand, the
effects of finite system sizes and of collective phenomena had not
been studied yet. Thus, for testing the survival and magnitude of
the effect in more realistic situations, we study the BBC when
mass-modification occurs in a finite sized, thermalized medium,
considering a non-relativistically expanding fireball with short
emission duration, and evaluating the width of the back-to-back
correlation function. We show that the BBC signal indeed survives
the expansion and flow effects, with sufficient magnitude to be
observed at RHIC. Some preliminary results are discussed here and
illustrated for particular cases.

%
Our analysis assumes the validity of local thermalization and
hydrodynamics up to the system freeze-out.
We also consider
${H}={H}_0-\int d{\bf x} d{\bf y} \phi({\bf x}) \delta M^2({\bf
x}-{\bf y})\phi({\bf y})$
as an effective in-medium Hamiltonian, where the first term
%
%
is the asymptotic (free) Hamiltonian in the rest frame of the
matter, and the second term describes the medium modifications.
The scalar field $\phi$ represents quasi-particles propagating
with momentum-dependent medium-modified mass $m_\star$, related to
the vacuum mass, $m$, by $
m_\star^2(|{\bf k}|)=m^2-\delta M^2(|{\bf k}|).
%
%
$ This implies that the dispersion relations in the vacuum and
in-medium are given, respectively, by $\omega_k^2 = m^2+{\bf k}^2
\nonumber$ and $\Omega_k^2 = m_\star^2+{\bf k}^2=\omega_k^2 -
\delta M^2(|{\bf k}|)$,
where $\Omega$ is the frequency of the in-medium mode with
momentum ${\bf k}$.

The annihilation (creation) operator, $b$ ($b^\dagger$), for the
in-medium, thermalized quasi-particles is related to the
annihilation (creation) operator, $a$ ($a^\dagger$), of the
asymptotic, observed quanta with momentum $k^\mu = (\omega_k,{\bf
k})$, by the Bogoliubov-Valatin transformation: 
$a_k=c_k b_k + s^*_{-k} b^\dagger_{-k} $ ( $ a^\dagger_k=c^*_k
b^\dagger_k + s_{-k} b_{-k}$ ),
%
where $c_k=\cosh(f_k)$ and $s_k=\sinh(f_k)$; $f_k=\frac{1}{2}
\log(\frac{\omega_k}{\Omega_k})$ is called {\sl squeezing
parameter}, since the Bogoliubov transformation creates squeezed
states from coherent ones.
In cases where the particle is its own anti-particle (for
$\pi^0$$\pi^0$ or $\phi$$\phi$ boson pairs, for instance), the
full correlation function is written as \be C_2({\mk_1},{\mk_2}) =
1 + \frac{|G_c({\mk_1},{\mk_2})|^2}{G_c({\mk_1},{\mk_1})
G_c({\mk_2},{\mk_2})} + \frac{|G_s({\mk_1},{\mk_2})
|^2}{G_c({\mk_1},{\mk_1}) G_c({\mk_2},{\mk_2}) }, \label{fullcorr}
\ee where the first two terms correspond to the HBT correlation,
and last term, represents this additional contribution to the
correlation function, i.e., the squeezing part.

For a hydrodynamical ensemble, both the chaotic and the squeezed
amplitudes, $G_c$ and $G_s$, respectively, are given by eqs. (22)
and (23) of Ref.~\cite{acg99}, which generalize  the formulae
derived by Makhlin and  Sinyukov\cite{MakSyniukov} to the case of
in-medium mass modification, as

%
\be G_c({\mk_1},{\mk_2}) = \int \frac{d^4\sigma_{\mu}(x)}{(2
\pi)^3} \, K^\mu_{1,2} \, e^{i \, q_{1,2} \cdot x} \,
\Bigl\{|c_{1,2}|^2 \, n_{1,2}(x) + |s_{-1,-2}|^2 \,
\left[n_{-1,-2}(x) + 1\right] \Bigr\}, \label{e:gc} \ee \be
G_s({\mk_1},{\mk_2})  = \int \frac{d^4\sigma_{\mu}(x)}{(2 \pi)^3}
\, K^\mu_{1,2} \, e^{2 \,i \, K_{1,2} \cdot x} \Bigl\{
s^*_{-1,2}\, c_{2,-1} \, n_{-1,2}(x) + c_{1,-2} \, s^*_{-2,1} \,
\left[n_{1,-2}(x) + 1\right] \Bigr\}. \nonumber\\
\label{e:gs} \ee

In Eq. (\ref{e:gc}) and (\ref{e:gs}),
$d\sigma^4_\mu(x)=d^3\Sigma_\mu(x,\tau)F(\tau)d\tau$ is the
product of the normal-oriented volume element depending
parametrically on $\tau$ (freeze-out hyper-surface parameter) and
on its invariant distribution, $F(\tau)$; $\sigma^\mu$ is the
hydrodynamical freeze-out surface. In Eq. (\ref{e:gc}), the pair
momentum difference and the pair
average momentum are given, respectively, by $%
q^{(*)\mu}_{i,j}(x) = k^{(*)\mu}_i(x) - k^{(*)\mu}_j(x)$, and
$K^{(*)\mu}_{i,j}(x) = \frac{1}{2} \left[k^{(*)\mu}_i(x) +
k^{(*)\mu}_j(x) \right]$, as in HBT;~ $c_{i,j}=\cosh(f_{i,j})$ and
$s_{i,j}=\sinh(f_{i,j})$, with
$ f_{i,j}(x)=\frac{1}{2}\log\left[\frac{K^{\mu}_{i,j}(x)\, u_\mu
(x)} {K^{*\nu}_{i,j}(x) \, u_\nu(x)}\right] = \frac{1}{2}
\log\left[\frac{\omega_{\mk_i}(x) + \omega_{\mk_j}(x)}
{\Omega_{\mk_i}(x) + \Omega_{\mk_j}(x)}\right] \equiv f_{\pm i,\pm
j}(x). $
We use a short-hand notation for the momenta, $\pm i, \pm j \equiv
k_{\pm i,\pm j}$, with $i,j=1,2$.

For studying the expansion of the system we adopt the
non-relativistic hydrodynamical model of Ref.~\cite{Csorgo:fg}. In
this model the fireball expands in a spherically symmetric manner
with a local flow vector given by the four-velocity $u^\mu =
\gamma \, (1, {\mathbf v})$, assumed to be non-relativistic, with
$\gamma = (1-{\mathbf v}^2)^{-1/2} \approx 1 +{\mathbf v}^2/2$,
where
${\mathbf v} = \langle u\rangle {\mathbf r}/R$,
being $\langle u\rangle$ and $R$ the mean expansion velocity and
the radius of the fireball, respectively.

In addition, we consider the Boltzmann limit of the Bose-Einstein
distribution for $n_k$, i.e., $n_{i,j}\sim \exp{[-(K^\mu_{i,j}
u_\mu - \mu(x))/T(x)]}$, and assume a time-dependent parametric
solution of the hydrodynamical equations, i.e.,
$\mu(x)/T(x)=\mu_0/T_0 - r^2/(2R^2)$, as in Ref. \cite
{{Csorgo:fg}}.
%
%
Furthermore, we consider a smeared freeze-out, for which
$\frac{\theta(\tau-\tau_0)}{\Delta \tau}e^{-(\tau-\tau_0)/\Delta
\tau}$, with short emission $\Delta t$. This more realistic
scenario has a dramatic effect on the Back-to-Back Correlation
function, as already showed in Ref.\cite{acg99}, by reducing
severely the signal's magnitude, even for a smearing of about
$\Delta \tau \sim 2$ fm/c. A clear illustration of the finite
emission duration as compared to the sudden freeze-out can be seen
in Ref.\cite{phkpc05,WPCF2005}.

For discussing finite-size effects, we distinguish between the
volume of the entire thermalized medium, denoted by $V$ (with
radius $R$), and the volume filled with mass-shifted quanta,
denoted by $V_s$ (with radius $R_s$). Naturally, $V_s \leq V$ in
the general case. In the derivation of the expressions for
$G_c(1,2)$ and $G_s(1,2)$, for simplicity, we consider that the
volumetric region where the mass $m_\star$ is significantly
modified is smooth and Gaussian in shape, i.e., we introduce a
three-dimensional Gaussian profile, $\sim \exp{\left[- {\bf
r}/(2R^2)\right]}$, for representing the system volume.

In the non-relativistic limit, the accounting for the squeezing
effects can be simplified for small mass shifts $(m_\star - m)/m
\ll 1$, such that the squeezing parameter can be written simply as
$ f(i,j,{\mathbf r}) \approx
\frac{1}{2}\log(m/m_\star)$. This limit is important, because in
this case the coordinate dependence enters the squeezing parameter
$f$ only through the possible position dependence of the
mass-shift which, in principle, could be calculated from thermal
field models in the local density approximation. Therefore, in an
approximation such that the position dependence of the in-medium
mass can be neglected, the $c(i,j)=c_0$ and $s(i,j)=s_0$ factors
can be removed from the integrands in Eq. (\ref{e:gc}) and
(\ref{e:gs}) and all what remains to be done are Fourier
transforms of Gaussian functions.


\begin{figure}[htb]
\begin{minipage}[t]{75mm}
\includegraphics[height=.33\textheight]{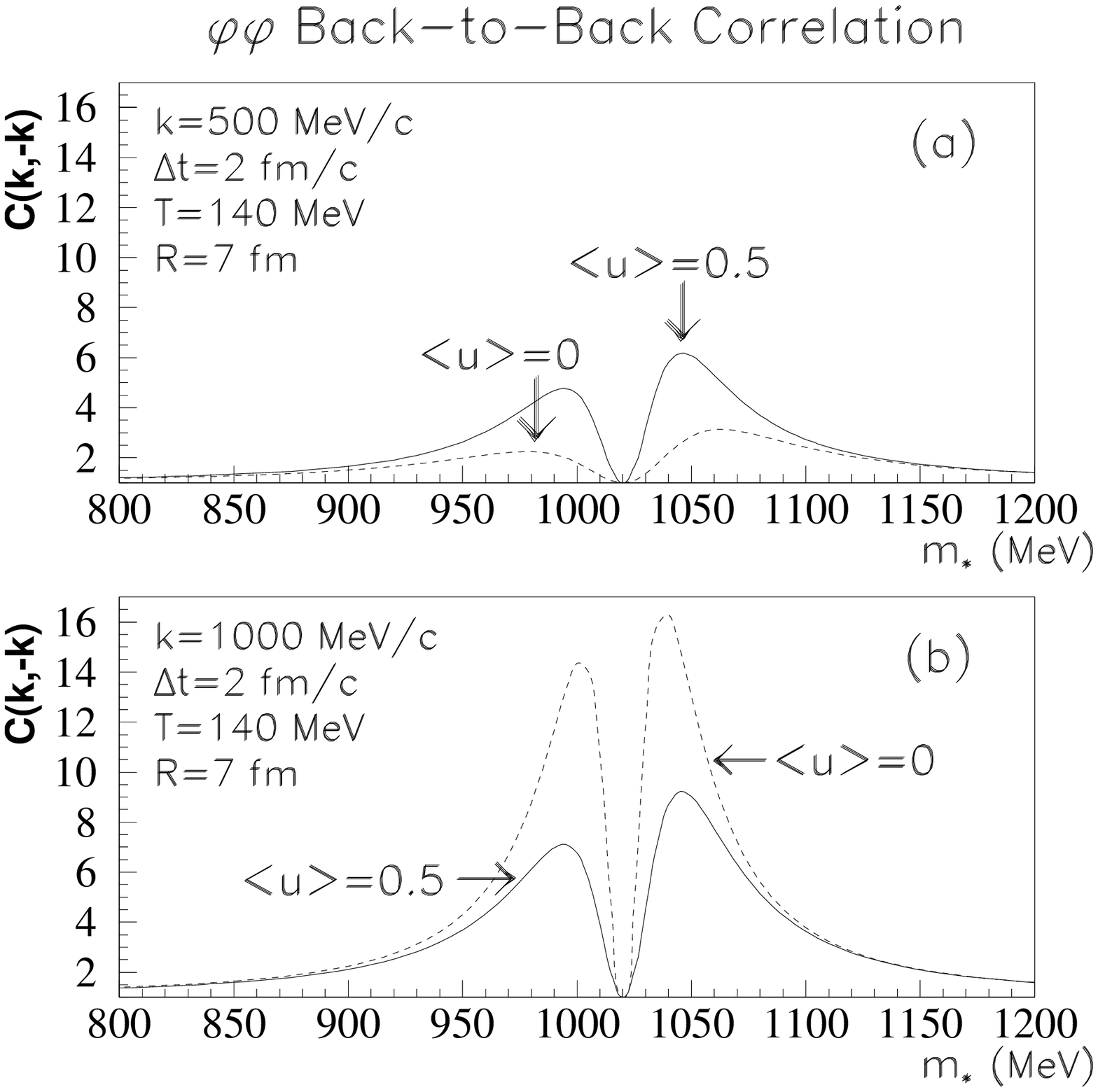}
\vskip-.5cm\caption{The maximal BBC is illustrated vs. $m_\star$,
when the mass-shift occurs in the entire system with radius $R$,
i.e., $V_s = V$, for two values of the momentum $k$ of the
back-to-back pair, with and without flow.}
\label{fig:largenenough}
\end{minipage}
\hspace{\fill}
\begin{minipage}[t]{75mm}
\includegraphics[height=.33\textheight]{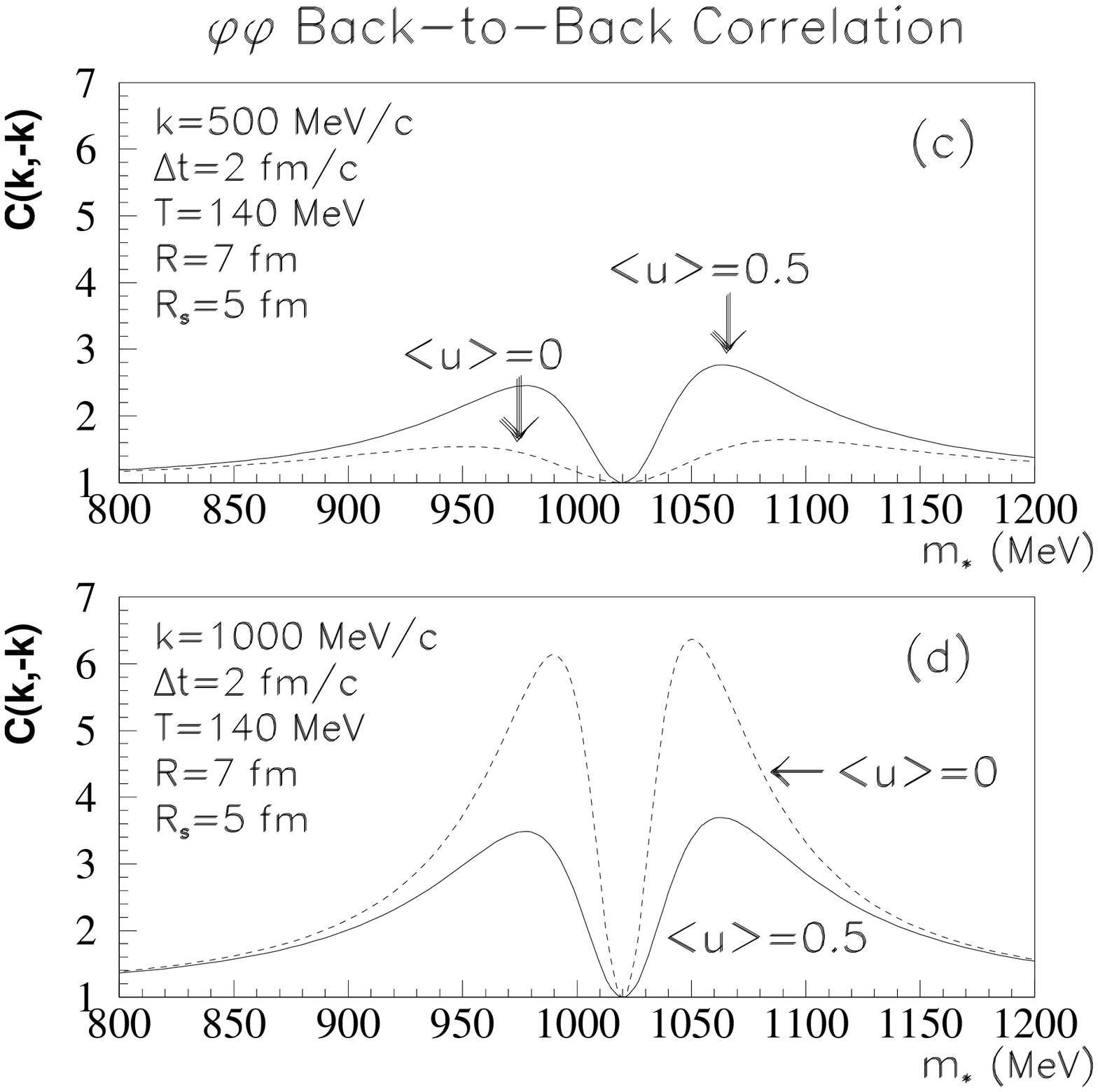}
\vskip-.5cm\caption{The maximal BBC vs. $m_\star$ is shown, for
mass-shift occurring only in a smaller portion of the system, with
radius $R_s<R$, i.e., $V_s < V$. All the other variables are
similar to the ones in Figure 1.} \label{fig:toosmall}
\end{minipage}
\end{figure}

\vskip-0.3cm In Figures 1 and 2, we illustrate some of the results
found in the non-relativistic approximation, in the particular
case of weak flow coupling. Figure 1 corresponds to the case in
which the mass-shift is extended over the entire system volume,
whereas in Figure 2, we show results for the squeezing occurring
in a smaller portion of the system region. In the plots, T stands
for the freeze-out temperature.  We see that the cases
corresponding to the absence of flow and to its inclusion produce
similar results within the limits of our illustration, and that
the strength of the squeezed BBC function is proportional to the
size of the mass-shifting region. However, depending on the value
of $k_1=-k_2=k$, there are noticeable differences. Being so, we
see that, for smaller values of $k$, the presence of flow seems to
slightly enhance the signal, whereas at large values of $k$, the
non-flow case wins. Nevertheless, the non-flow case grows faster
with increasing $k$.


In summary, our main goal on presenting these new results here was
to revive the discussion on the search of the squeezed BBC. For
fulfilling this purpose, we estimated the strength of the squeezed
BBC signal in a more realistic situation, considering the
mass-shifting in a finite region, and the particle emission
occurring during a short interval. We also considered that the
system expands non-relativistically and analyze the simplified
situation of weak flow dependence of the squeezed BBC. For
illustrating the effects, we considered $\phi\phi$ pairs. We
showed in Figures 1 and 2  the back-to-back correlation function
versus the in-medium shifted mass, $m_\star$, with pronounced
maxima around $m \approx m_\star$. We also saw that both the
non-flow and the flow cases produced similar results, and that the
BBC magnitude increases proportionally to the size of the
mass-shifting region. However, for reducing the BBC magnitude, we
saw that the effect of decreasing the system size is far less
significant than the sensitivity to the spread in the time
emission interval. Our main conclusion, nevertheless, is that in
any of the two cases discussed above, a sizeable strength of the
squeezed BBC signal could be seen, making it a promising effect to
be searched for experimentally at RHIC.

Naturally, in a more refined calculation, it would be mandatory to
introduce a model-based mass-shift. On top of that, it would also
be essential to perform more realistic calculations with flow, in
a less constrained kinematical region, while simultaneously
searching for those windows which could optimize the observation
of the squeezed BBC signal. Also, an estimate of the shape and
width of the BBC around the direction $k_1=-k_2=k$ should be
performed.
Finally, for being able to make predictions closer to the
experimental conditions, it will be extremely important to obtain
some feed-back on the experimental acceptance, conditions, and
restrictions that could finally lead to the BBC discovery.


\begin{thebibliography}{9}

\bibitem{acg99} M. Asakawa, T. Cs\"org\H o and M. Gyulassy, Phys.
Rev. Lett. {\bf 83} (1999) 4013.
\bibitem{pchkp01} P. K. Panda, T. Cs\"org\H o, Y. Hama, G. Krein and Sandra S.
Padula, Phys. Lett {\bf B512} (2001) 49.
\bibitem{MakSyniukov} A. Makhlin and Yu. Sinyukov,  Yad. Phys. {\bf 46}
(1987) 637; Yu. Sinyukov, Nucl. Phys. {\bf A566} (1994) 589c.
\bibitem{Csorgo:fg} T.~Cs\"org\H{o}, B.~L\"orstad and J.~Zim\'anyi,
Phys.\ Lett.\ {\bf  B338} (1994) 134; P. Csizmadia, T.
Cs\"org\H{o} and B. Luk\'acs, Phys.Lett. {\bf  B443} (1998) 21.
\bibitem{phkpc05} Sandra S. Padula, Y. Hama, G. Krein, P. K.
Panda, and T. Cs\"org\H{o}, in preparation.
\bibitem{WPCF2005} Sandra S. Padula, Y. Hama, G. Krein, P. K.
Panda, and T. Cs\"org\H{o}, Proc. Workshop on Particle
Correlations and Femtoscopy (WPCF 2005), in press [nucl-th/0510068].

\end{thebibliography}
\end{document}